# Tracking dynamic flow: Decoding flow fluctuations through performance in a fine motor control task

Bohao Tian, Shijun Zhang, Sirui Chen, Yuru Zhang, *Senior Member, IEEE*, Kaiping Peng, Hongxing Zhang and Dangxiao Wang, *Senior Member, IEEE*

*Abstract*—**Flow, an optimal mental state merging action and awareness, significantly impacts our emotion, performance, and well-being. However, capturing its swift fluctuations on a fine timescale is challenging due to the sparsity of the existing flow detecting tools. Here we present a fine fingertip force control (F³C) task to induce flow, wherein the task challenge is set at a compatible level with personal skill, and to quantitatively track the flow state variations from synchronous motor control performance. We extract eight performance metrics from fingertip force sequence and reveal their significant differences under distinct flow states. Further, we built a learning-based flow decoder that aims to predict the continuous flow intensity during the user experiment through the selected performance metrics, taking the self-reported flow as the label. Cross-validation shows that the predicted flow intensity reaches significant correlation with the self-reported flow intensity (*r*=0.81). Based on the decoding results, we observe rapid oscillations in flow fluctuations during the intervals between sparse self-reporting probes. This study showcases the feasibility of tracking intrinsic flow variations with high temporal resolution using task performance measures and may serve as foundation for future work aiming to take advantage of flow's dynamics to enhance performance and positive emotions.**

## I. INTRODUCTION

**M**any are familiar with entering a mental "zone" characterized by intense concentration and self-absorption when engaging in compelling tasks [1]. In this state, our performance reaches its peak, but the state itself is temporary and may abruptly disappear after a period. This familiar state is known as flow, in which an individual is completely absorbed in an activity without reflective self-consciousness but with a deep sense of control. Flow plays a pivotal role in promoting positive emotion [2], enhancing learning effect [3], and enhance interpersonal relationships [4]. However, studies focusing on the dynamics of flow reveal its unpredictability and discontinuity, often experienced as abrupt transitions into and out of the state [5]. Capturing the intrinsic fluctuations of flow holds the potential to unveil the mechanism behind its highly transient nature and effectively harness its dynamic properties. For instance, enterprises can optimize the work rhythm or tailor task difficulty in real-time, leveraging the dynamic characteristics of flow experience. Such adaptation aims to enhance employee engagement and well-being.

Regrettably, current flow detection and measurement tools suffer from limitations in temporal resolution, hindering the comprehensive exploration of its temporal dynamics. Experience Sampling Methods (ESM) have been widely employed to investigate the dynamic patterns of flow [6], providing valuable insights into its temporal variations and uncovering phenomena such as chaos and hysteresis [7][8][9][10]. Whereas, the limited temporal density of sampling in ESM studies hampers their ability to capture the rapid fluctuations of flow occurring within minutes. This limitation is particularly problematic for studying and inducing flow in the tasks with short durations and rapidly changing contexts, such as aerial navigations, surgical operations, or sports like skiing and gliding [11]. Real-time detection of flow would set the initial reference for the closed-loop human-computer interaction system to manipulate the flow state and to enhance task engagement. To achieve this goal, it is imperative to fill the gaps between ESM samplings during spare intervals. Therefore, this work aims to decode the fluctuations of flow state over time and explore its dynamics on a second timescale.

However, the subjective, complex, and multifaceted nature of flow experience poses several challenges when attempting to decode it on a fine timescale. Numerous studies have emphasized that flow is composed of several mental components, including concentration, autotelic reward, cognitive control, etc. [12][13]. Firstly, the subjectivity of flow necessitates the predominant use of self-reporting as the measurement approach, like the Flow Short Scale (FSS). The sampling sparsity of such measurements limits the timescale to the sub-hour level [14]. While some signals, such as Electroencephalogram [15][16][17], facial expression [18], physiological recording [19][20][21], etc., offer the potential for higher temporal resolution in flow measurement, but flow state over a set period is typically calibrated based on self-

Corrsponding author: Dangxiao Wang (e-mail: hapticwang@buaa.edu.cn) and Hongxing zhang (e-mail: zhanghx08@126.com).

Bohao Tian, Shijun Zhang, Sirui chen, Yuru Zhang, and Dangxiao Wang are with the State Key Laboratory of Virtual Reality Technology and Systems, Beihang University, 100191 Beijing, China (e-mail: tianbohao @buaa.edu.cn, zhangshiji0103@qq.com, 20374152@buaa.edu.cn, {yuru and hapticwang}@buaa.edu.cn).

Kaiping Peng is with the Department of Psychology, Tsinghua University, 100084 Beijing, China (e-mail: pengkp@mail.tsinghua.edu.cn)

Hongxing Zhang is with the State Key Laboratory of Medical Proteomics, Beijing Proteome Research Center, National Center for Protein Sciences (Beijing), Beijing Institute of Life Omics, 102206 Beijing, China

Dangxiao Wang are also with the Beijing Advanced Innovation Center for Biomedical Engineering, Beihang University, 100191 Beijing, China and the Peng Cheng Laboratory, 518055 Shenzhen, China.

reports or task challenge type, making it difficult to capture flow variations within minutes. Secondly, inducing the elusive flow experience under lab settings has proven challenging, requiring the manipulation of key factors such as optimal challenge, clarity of goals and feedback, and minimizing distractions [22]. Lastly, establishing and validating the models to describe flow is difficult due to its subjective and intricate nature, particularly regarding how the human behavior parameters are associated with flow [23].

Here, to break the limit of the timescale of flow detections due to the sparsity of self-reporting, we propose a novel decoding approach that utilizes real-time task performance to track the flow state variations. The cornerstone of flow decoding lies in inducing it and monitoring its variations over time using explicit performance metrics, which necessitates a reliable task paradigm. The balance between challenge and skill has long been recognized as a core requirement for inducing flow, and previous research has employed computer games like Tetris [24][25], mental arithmetic [26][27] and simulated vehicle piloting tasks [28][29][30] to empirically induce flow. These experimental methodologies have advanced the study of flow from the observation on the people's daily life to the rigorous research under lab settings. However, in these approaches, the match between challenge and skill is typically established by manually setting task difficulty to moderate, often overlooking individual differences in skill. Furthermore, to pursue high immersion and prevent boredom, these tasks lacked controlling over irrelevant factors and explicit performance indicators, limiting the quantitative tracking of flow. To address these limitations, we introduce a fine fingertip force control ($F^3C$) task that allows precise manipulation of task difficulty to align with an individual's force control skill. We extract several metrics from the force output sequence to represent performance variations. Given that the task comprises consecutive trials of 3-second duration, the performance metrics enable the tracking of flow on a second timescale.

To decode the flow variations via task performance metrics, it is crucial to establish the relationship between flow and performance. Previous studies have predominantly approached this by correlating self-reported flow experience with task performance of each subject [28][31][32], or proposing theories based on the components of flow [33]. These studies often treated the flow and task performance as static and rarely incorporate their variations over time. Against this backdrop, we need to expand upon the static flow model by incorporating performance variations. According to the classical theory of flow [1], for a specific task, the flow state can be entered only if the challenge is fixed at the personalized skill level. Under that condition, the user's mental state is presumed to fluctuate between the in-flow state and the out-flow state dominated by the intrinsic dynamics of the flow. Previous studies introduced the concept of flow intensity to differentiate between varying degrees of the flow experience [34]. It can be posited that flow intensity fluctuates over time when the challenge matches the skill, as illustrated in Fig. 1

(a) and (b). Building upon this notion, we propose two hypotheses regarding the relationship between flow and task performance:

*H1*: The task performance may be influenced by the flow state fluctuation and variates along with the flow over time.

*H2*: If *H1* was validated, the flow state variations may be predicted by the synchronous task performance metrics.

In this study, we collect the data of self-reported flow and task performance in the user experiment. We validate *H1* by extracting eight performance metrics from the fingertip force sequence and comparing these metrics under distinct self-reported flow states. The metrics extraction is supported by a model describing the process of hand-eye coordinated force control during the $F^3C$ task. Subsequently, to validate *H2*, we endeavored to decode flow variations by machine learning approach using these eight performance metrics. we constructed a linear regression-based decoder due to its explainable nature. However, this may bring the potential risk of overfitting caused by the sparsity of the self-reporting of flow. To address this, we opt to select those metrics with superior predictive capability for each participant, instead of using all the metrics. Furthermore, we investigate the dynamics underlying the flow fluctuations on a second timescale based on the cross-validated flow decoder. Our findings reveal the presence of swift oscillations of flow variations within the intervals between self-reporting probes and offers a dynamic perspective on the relationship between flow and task performance.

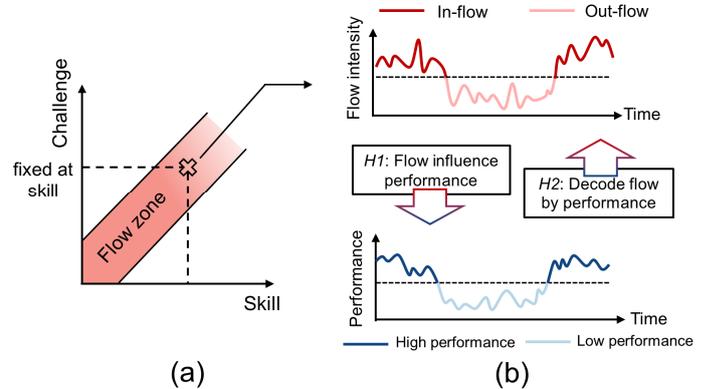

**Fig. 1**. Schematic of the two hypotheses underlying the relationship between flow and the task performance. (a) For a specific task, the flow zone can be entered only if the challenge matches the personal skill level. (b) The task performance may be influenced by the flow fluctuation and the flow intensity variation may be decoded by the task performance over time.

The real-time identification of an individual experiencing flow or not holds significant implications across various activities. By presenting a quantitative detection approach for flow fluctuations on a second timescale, our research not only broadens the computational methodology for exploring the intrinsic dynamics of flow with high temporal resolution but also establishes a cornerstone for the development of closed-loop human-computer interaction systems. Such systems hold promise in monitoring and regulating flow experience within real-world dynamic contexts, offering potential benefits in pursuing peak performance and fostering positive emotions.

## II. Materials and Methods

### 2.1. Fine Fingertip force control task

The antecedents of the flow state comprise clear goals, unambiguous feedback, and a perfect match between skill and challenge [1]. Herein, we designed a F³C task to induce the flow experience: participants were instructed to precisely control their fingertip force within a specified target range by pressing the force transducer according to the visual presentation. The immersive virtual reality (VR) environment, provided by a head-mounted display device (Oculus Rift DK2), aimed to take over the sensory and minimize distractions, helping the participants concentrated on the task. The design principles were depicted in Fig. 2(a).

As illustrated in Fig. 2(b), we constructed a virtual disk and a cylinder with adjustable thickness within a virtual scenery using the Unity 3D engine. At the beginning of each trial, the virtual disk was positioned at the bottom of the user's view, while the virtual cylinder remained fixed at the center. As the participant pressed the force transducer (Arizon Technologies, mediated by the homemade sampling board, with the measuring error of 0.02%) using the right index finger, the virtual disk synchronously rose to a height proportional to the vertical pressing force. Participants were forbidden from pressing the transducer prior to the appearance of the virtual cylinder at the trial onset, ensuring that the pressing force sequence always starts at zero in each trial. Only by continuously maintaining the force $F_t$ (denoted by the disk's height) in the target range $\Delta F$ (denoted by the thickness of cylinder) for a duration of over 500ms in a single trial with an upper time limit of 3s will the participant succeed in the current trial. The green and red visual feedback denoting success and failure respectively would be presented after each trial. Participants were instructed to stabilize the force as much as possible during each trial and to strive for a higher number of successful trials. As depicted in Fig. 2 (b), the center of the grey cylinder corresponded to a vertical pressing force with the magnitude of 1N. The thickness of the cylinder, (i.e., the target range $\Delta F$) served as an indicator of the task difficulty of the current trial. Smaller thickness corresponded to more challenging trials, requiring participants to exert greater effort to maintain the pressing force within the target range.

To account for individual variations in force control skill, we developed an adaptive procedure aimed at accurately measuring the skill for force control and precisely matching the task challenge to skill level. Drawing inspiration from the staircase approach used in psychophysiology [35], our method involved adjusting the task difficulty based on participants' performance until a convergence point was reached. This convergence point was considered as the measured skill of fingertip force control (see Appendix A and Supplementary Fig. 1). During the main experimental sessions, the task difficulty $\Delta F$ remained fixed at the measured skill, ensuring an optimal match between the challenge and skill.

### 2.2. Force control process and performance metrics

In the first hypothesis (*H1*), we assumed that the task performance would be influenced by the flow intensity variations over time. In the context of the closed-loop force control process, this can be theoretically interpreted through the lens of the closed-loop force control process. As depicted in Fig. 2(b), during the task, participants' current fingertip force was translated into the height of the disk, and the discrepancy between the disk and cylinder served as the reference to made decision to alter the fingertip force. Then, the force was adjusted based on the motor command. It is believed that being in a flow state can expedite this force control loop, consequently influencing task performance. We assumed that the flow intensity affected force output behavior, mediated by two critical parameters in the force control process: the updating period $\Delta T$ of the loop and the minimum modulation step of force $\delta F$. As flow intensity increases, the force control loop would be facilitated to work, leading to a decrease in $\Delta T$ and $\delta F$. While the task performance would variate along with the decreasing of $\Delta T$ and $\delta F$.

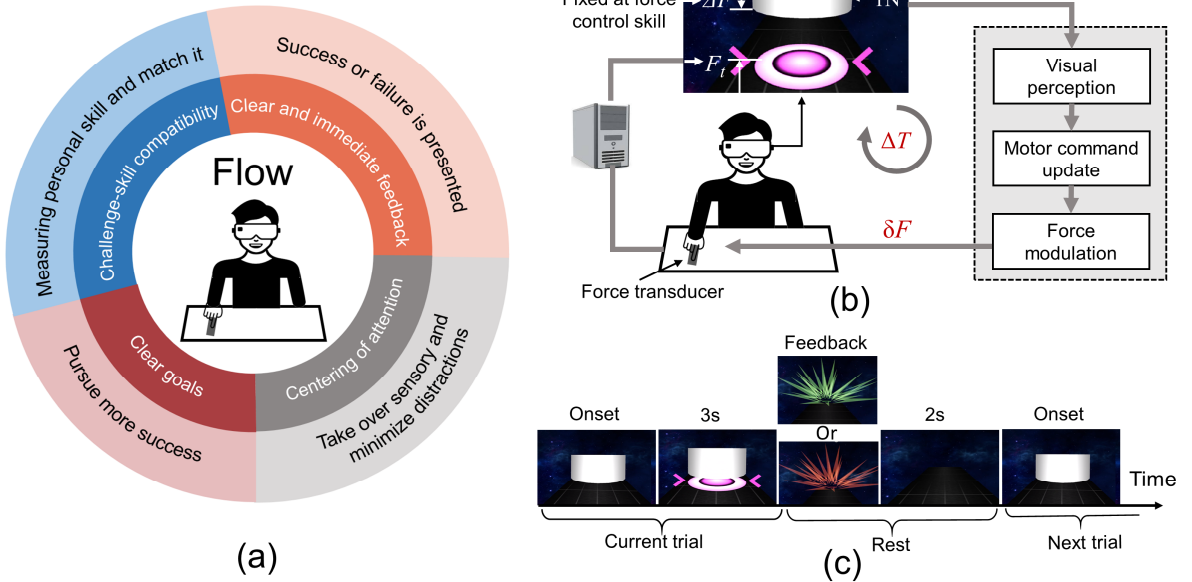

**Fig. 2.** The F³C task to induce the flow. (a) Schematic of the design principles. The F³C task is designed to follow the main antecedents and components of flow. The inner circle represents the conditions of the flow and the outer circle represents the design elements of the task. (b) The details of the F³C task. During the task, the participant is instructed to press the force transducer using the right index finger to control the virtual disk (the height denotes the pressing force $F_t$) maintaining within the target range (denotes the force range $\Delta F$) as much as possible. (c) The procedure of the F³C task for a single trial, encompassing pressing duration of 3s and resting duration of 2s.

To explicitly characterize the task performance of each trial, we defined eight metrics from the fingertip force output sequence within a single trial, as follows and demonstrated in Fig. 3(a):

*1) Reaction time.* The reaction time is defined as the duration from the trial onset to the timepoint when the force transducer detected the pressing force, which could be represented by $t_1$ in Fig. 3(a).

*2) Arriving Time.* The arriving time is defined as the duration from beginning of the pressing to the timepoint when the fingertip force $F_t$ reach the target range at the first time, i.e., $t_2$ in Fig. 3(a).

*3) Completing time.* The completing time is defined as the duration from the trial onset to the timepoint the participant met the success condition before the trial is finished. The success condition is defined as a duration exceeding 500ms where pressing force was continuously in the target range (C.T. in Fig. 3(a)).

*4) In-range time.* The in-range time indicates the cumulative duration of which the pressing force was in the target range before the trial is finished, which could be represented as the summation of $t_3$, $t_4$, and $t_5$ in Fig. 3(a). Even though $t_3$ and $t_5$ are not to exceed 500ms.

*5) Force overshooting.* The force overshooting $F_{OS}$ was defined as the peak of the fingertip force relative to the target force, as follows:

$$F_{OS} = \left| F_{max} - F_T \right| / F_T \tag{1}$$

where $F_{max}$ denotes the peak force during the single trial and $F_T$ denotes the target force (i.e., 1N).

*6) Average deviation.* The average deviation $F_{AD}$ was defined as the averaged discrepancy between the current force and the target force at each sample, as follows:

$$F_{AD} = \sum\nolimits_{i=1}^{N_S} \left| F_i - F_T \right| / N_S \tag{2}$$

where $N_S$ represented the number of the force samples, and $F_i$ denoted the force at the current sample.

*7) Average adjusting rate.* The average adjusting rate $F_{ADR}$ was defined as the average change of the fingertip force at two adjacent samples, as follows:

$$F_{ADR} = \frac{\sum_{i=2}^{N_S} \left| F_i - F_{i-1} \right|}{(N_S - 1)\Delta t} \tag{3}$$

where $F_i$ and $F_{i-1}$ denoted the fingertip force at the current and the last sample respectively, $\Delta t$ was the sampling period (i.e., 0.001s).

*8) Success rate.* For a single trial, the success or failure was marked as 1 and 0 respectively. Over several trials, the success rate was defined as the number of successful trials divided by the number of total trials.

These metrics were specifically designed to capture the influence of $\Delta T$ and $\delta F$ and thereby distinguish different flow states. The influences on these metrics were supported by the simulation performed following the mathematical model for the fingertip force control process (see Appendix B), as demonstrated in Supplementary Fig. 2 and 3. To ascertain the ground truth of flow intensities, we placed twelve self-reporting probes throughout the user experiment (see Section 3.2). The hypothesis stating that the task performance was influenced by the flow intensity variations can be validated by the self-reported flow intensity and the associated performance metrics of each flow probe in the experiment.

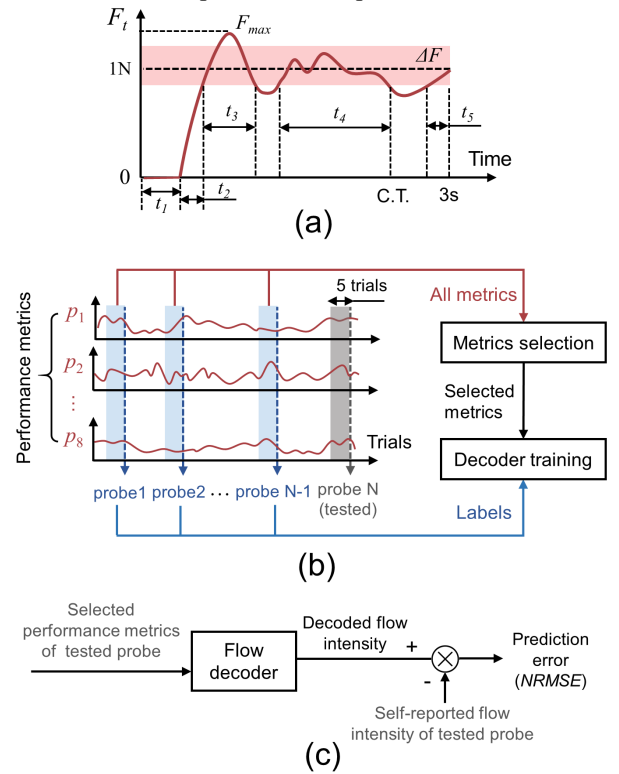

**Fig. 3.** Decoding framework of flow. (a) Schematic of the fingertip force output sequence within a single trial. The light red band shows the target range of force control (i.e., $\Delta F$). $t_1$ to $t_4$ denoted the temporal durations within the trial. C.T. stands for the completing time. (b) Construction of the leave-one-out cross-validated flow decoder. For each participant, a flow probe (as an example, probe N) is left out as the tested probe to be predicted. The remaining flow probes and associated performance metrics are utilized to train the decoder. (c) The trained decoder is used to predict the tested flow probe (probe N) based on its associated selected performance metrics.

### 2.3. Construction of the flow decoder

In the second hypothesis (*H2*), we assumed the flow intensity could be decoded by the synchronous task performance. For the F³C task, the task performance was characterized by the above eight metrics and the flow intensities were labelled by the self-reported results. For simplicity, the flow decoder can be represented by the linear mapping between the predicted flow intensity $\hat{I}_t$ and the synchronous performance metrics, as follows:

$$\hat{I}_t = \boldsymbol{T}^* \boldsymbol{P}_t + I_0^* \qquad (4)$$

where $\boldsymbol{P}$ was the performance metrics vector, $\boldsymbol{T}^*$ and $I_0^*$ were the coefficients, which were determined within the flow decoder training. To select the metrics with better predictive capability while mitigating the risk of overfitting, we set a maximum limit of the decoding-used metrics as four and conduct the metrics selection process to obtain the metrics vector $\boldsymbol{P}$, instead of using all the eight metrics.

Let $\boldsymbol{\Omega}$ represent the full set of metrics vectors, $\boldsymbol{P}=[p_{a1},$ $p_{a2}, \ldots, p_{an}]^T \in \boldsymbol{\Omega}$, where $p$ came from the eight performance metrics we defined (i.e., $1 \leqslant a_1 \leqslant a_n \leqslant 8$). Supposing $n$ metrics were selected for decoding ($n \leqslant 4$), then $\boldsymbol{P} \in R^{n \times l}$ and $\boldsymbol{T}^* = [t_1, t_2\ldots, t_n] \in R^{l \times n}$. For specific participant, we picked out every $\boldsymbol{P}$ within $\boldsymbol{\Omega}$ to train the flow decoder, ultimately identifying the $\boldsymbol{P}$ associated with the optimal decoding performance as the designated decoding metrics vector. This metric selection procedure effectively ensured the number of flow probes surpassing three times the number of decoding features for each subject. Additionally, this process facilitated an understanding of which metrics exhibited greater diagnostic potential within the flow decoding context.

The flow decoding performance was evaluated by the leave-one-out cross-validation, as illustrated in Fig. 3(b). For twelve self-reporting flow probes of each participant, one probe was left out as the tested probe to be predicted. The remaining probes and associated performance metrics were utilized to train the decoder. We calculated each metric from the force output sequence of the previous five trials relative to the insertion timepoint of each probe (the number of trials was accordance with the flow probe questions) and averaged them across five trials as this metric associated with that probe. The decoder training process was to calculate $\boldsymbol{T}^*$ and $I_0^*$ to optimize the fitting between the predicted and the self-reported flow intensity, as follows:

$$\left(\boldsymbol{T}^*, I_0^*\right) = \underset{\boldsymbol{T}, I_0}{\operatorname{argmin}} \sum_{k=1}^{N-1} \left\| I^{(k)} - \boldsymbol{T}\boldsymbol{P}^{(k)} - I_0 \right\|_2 \qquad (5)$$

where $N$ denoted the number of all flow probes (i.e., twelve), $I^{(k)}$ and $\boldsymbol{P}^{(k)}$ denoted the self-reported flow intensity and performance metrics vector associated with the $k$-th flow probe, and $\boldsymbol{P} \in \boldsymbol{\Omega}$.

Once calculate $\boldsymbol{T}^*$ and $I_0^*$ by solving the optimization, the prediction of the tested flow probe could be obtained using the Eq. (4), and this process was repeated twelve times. By iteratively searching each metrics vector $\boldsymbol{P}$, we are capable of identifying the most optimal combination of performance metrics to effectively decode the flow.

### 2.4. Assessments of the flow decoder

The prediction error of flow decoding was determined by the discrepancy between the self-reported flow intensity of the tested probes and their prediction in each validation loop, as illustrated in Fig. 3 (c). We utilized the normalized root mean square error (*NRMSE*), a widely accepted index to represent prediction error. It was defined as:

$$NRMSE = \sqrt{\frac{\sum_{j=1}^{N} \left( \hat{I}^{(j)} - I^{(j)} \right)^2}{\sum_{j=1}^{N} \left( I^{(j)} \right)^2}} \qquad (6)$$

where $I^{(j)}$ and $\hat{I}^{(j)}$ denoted the self-reported flow intensity and the prediction result of the $j$-th probe respectively. The smaller *NRMSE* represented the higher decoding accuracy. The correlation between the self-reported and the predicted flow intensities of the twelve probes can be used to assess the decoding performance for a certain participant.

Besides the correlation analysis, two statistical tests were employed to assess the flow decoder. (i) In the random-test, for each participant, we generated 1,000 sets of random sequences for each participant using the numbers drawn from the same range as self-reported flow intensity and place these at each probe. Cross-validated decoding was then performed repeatedly, using the same performance metrics, for each set of random flow labels. Consequently, we obtained a distribution of 1,000 cross-validated prediction errors following Eq. (6) for each subject. The random-test *P*-value was defined as the probability of obtaining lower cross-validated prediction error with random flow labels compared to true self-reported flow labels. (ii) In the permutation-test, we randomly permuted the time indices of the flow probes 1,000 times for each participant and repeated the procedure in the random-test to obtain the permuted-test *P*-values. The random-test was considered the primary criterion for determining significance, while the permuted-test was conducted to demonstrate the robustness of the decoder [36].

## III. USER STUDY

### 3.1. Participants

We recruited 32 right-handed adults from Beihang University [age $26.2 \pm 4.4$ y (mean $\pm$ SD); eleven women]. Only subjects who reported normal vision or vision that was corrected to normal with contact lenses were included. We excluded volunteers who reported using medication that might influence the experiment. Participants were compensated with $10 per hour. The experiment was approved by the Biological and Medical Ethics Committee of Beihang University (No. BM20230123) and was performed in accordance with the Declaration of Helsinki. Written informed consent was obtained from all participants before the experiment.

Two participants (both men) were excluded because their success rate was too high (exceeding 90%). We speculated they did not understand the pre-test task instructions on the skill measurement, resulting in the subsequent task that was too easy for their skill. They were excluded because the excessive successful rate meant the condition of flow occurrence, i.e., the challenge is as close to the skill as possible, cannot be ensured. Six participants (two women) were excluded for their self-reported results of flow probes. If the range of the self-reported flow intensity is too narrow, it is not sufficiently effective to be the ground truth for the decoding of flow variations. Hence, the participants whose range of self-reported flow intensity is less than 1 (for the seven-level Likert scale) were excluded. In the end, 24 participants [age $26.2 \pm 4.4$ y (mean $\pm$ SD); nine women] were left for subsequent analyses, guaranteeing a match between challenge and skill and effectiveness of self-reporting, both of which are essential for flow decoding.

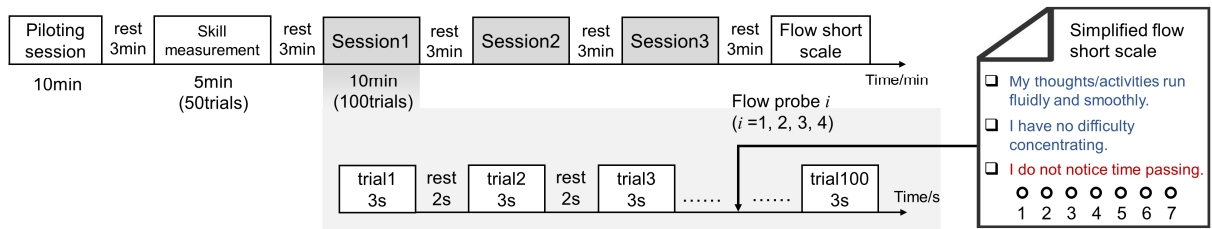

**Fig. 4.** The experimental procedure and the self-reporting flow probe. Twelve flow probes are interspersed throughout the experiment to inquire the participants' current flow state. The blue and red questions represent the dimension of fluency and absorption respectively. The instruction for the participant when him/her encounters the flow probe is: "*Please choose the answer from one to seven based on your thoughts and feelings in the previous five or so trials. There are no right or wrong answers*".

### 3.2. Procedure and self-reporting flow measurements

The experiment consisted of several components: a skill measurement session to assess the participant's force control skill, three sessions of the main experiment with a fixed task difficulty matching the measured skill, and a full-version Flow Short Scale (FSS) [14] to assess flow experience induced by the $F^3C$ task during the experiment at the end, as illustrated in Fig. 4.

Prior to the experiment, the participants were required to practice the $F^3C$ task till they adequately constructed the mapping relationship from fingertip motor command to the visual feedback provided in VR. The skill measurement session encompassed 50 trials, following the adaptive procedure (see Appendix A). Each main session comprised 100 trials, with 3-minute intervals between adjacent sessions to allow the participants to rest. Thus, each participant performed the $F^3C$ task for a total of 300 trials. The force output data during each trial were recorded at a sampling frequency of 1kHz for subsequent analysis.

During the experiment, it was essential to measure the self-reported flow intensity at different time as the ground truth for tracking flow variations. However, self-reporting can disrupt the continuity of the task and potentially interrupt the flow experience. Therefore, it was crucial to limit the number of self-reporting timepoints, also known as flow probes. A total of twelve flow probes were used, with four probes inserted in each session. The interval between two adjacent probes was set to a minimum of twelve trials to avoid excessive insertion of probes within a short time period. When a flow probe was inserted, participant would encounter a pop-up dialog box in the virtual scenery, inquiring about their current flow state.

FSS consisted of ten questions that can be categorized into two dimensions: the fluency of performance (six questions) and the absorption by activity (four questions) [14]. It employed a seven-level Likert scale to assess subjective flow experience (1=not at all; 4=partly; 7=strongly agree). The average of the participants' responses to the ten questions served as the assessment result. To minimize the interruption, we carefully selected three questions from the FSS that still encompassed both dimensions to compose the flow probe. The selection was based on interviews conducted with participants from a preliminary study, wherein we inquired about their experimental experiences. The content of the flow probe was illustrated in Fig. 4. Among the three selected questions, the first and second questions (blue in Fig. 4) pertain to the dimension of fluency, while the third question related to the dimension of absorption by activity (red in Fig. 4). The average of the participants' responses to these three questions served as the self-reported flow intensity for the current probe.

### 3.3. Data analysis and statistics

All the behavioral data and self-reported data for analysis were collected in the user experiment. The two-tailed independent samples $t$-tests were conducted on the comparison of the eight performance metrics associated with the probes of different flow states. The two-tailed paired samples $t$-tests were conducted to compare the prediction error among flow decoding, random-test, and permutated-test, as well as to compare the prediction error among the decoding of flow, fluency, and absorption.

After construct and cross-validate the flow decoder, the power spectral density (PSD) analysis was utilized to quantify the timescale of the fluctuation of the decoded flow intensity time series. The PSD provided insights into the relative prominence of different timescales, which were defined as the inverse of frequency, within the decoded flow intensity time series. PSD analysis was conducted by the Welch approach, which was implemented by the function *pwelch* in MATLAB. All the statistics in this work were implemented using the SPSS (version 25.0) and MATLAB (version 2018b) and demonstrated after corrected with false discovery rate using the Benjamini–Hochberg approach.

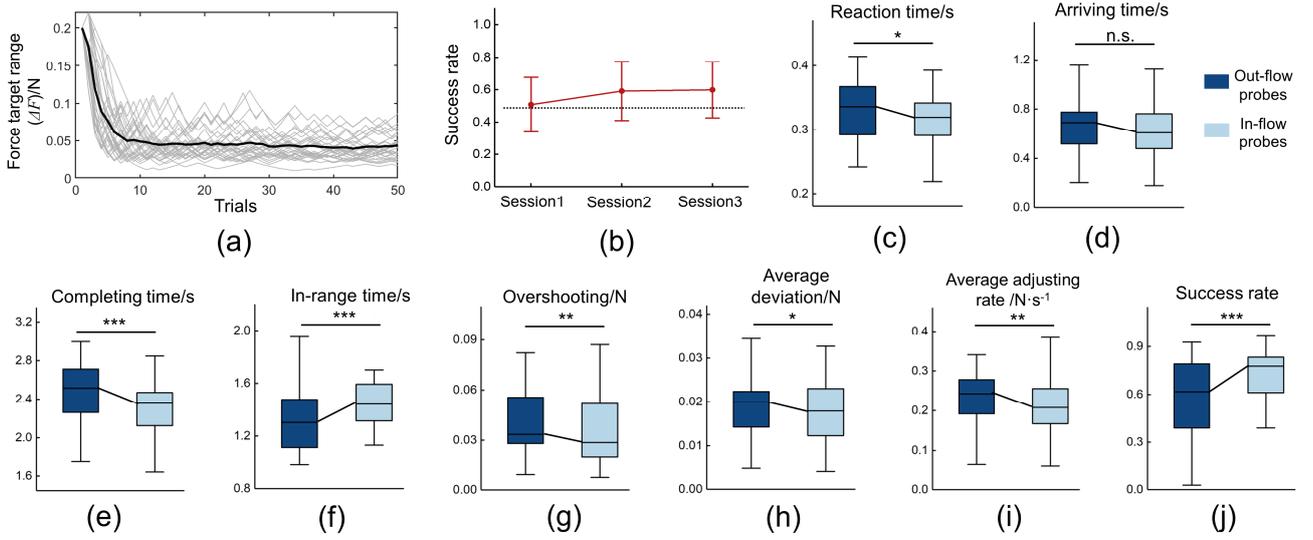

**Fig. 5.** The behavioral measures. (a) The task difficulty variations during the measurement session. The task difficulty shows the tendency of decreasing and converging. Each grey line denotes each participant and the bold black line denotes the average variation across all participants. (b) The success rate during the three main experimental sessions. The error bars denote the standard deviation across subjects and the dot line denotes the success rate of 0.5. (c) to (j) demonstrate the eight performance metrics extracted from the force output sequence associated with the in-flow and out-flow probes across participants, compared by paired $t$-tests ($N$=24). Hinges of boxplots represent the first and third quartile and whiskers span from the smallest to the largest value of the data. $*P<0.05$, $**P<0.01$, $***P<0.001$, and n.s. stands for no significance.

## IV. RESULTS

### 4.1. Flow induction

Following the experimental procedure, the skill of fingertip force control (represented by target force range $\Delta F$) of each participant was measured before the three main sessions. The task difficulty variation of each participant during the measurement session was demonstrated in Fig. 5(a). The average measured skill across all participants was $0.040 \pm 0.013$N (mean ± SD). The success rates of each experimental session were demonstrated in Fig. 5(b), in line with our hypothesis that the probability of success would be 0.5 when task difficulty matched the personal force control skill (see Appendix A). The average score across all participants on the full-version FSS, which was employed in the end of the experiment, was $5.48 \pm 0.71$ (mean ± SD, averaging ten questions within the seven-level scale), suggesting that the F³C task effectively elicited flow experience.

### 4.2. Performance metrics

The eight performance metrics were extracted to validate the relationship between the task performance and flow. For each participant, the median value of the self-reported flow intensities of all probes served as the criteria to classify the probes into in-flow and out-flow probes. As demonstrated in Fig. 5 (c) to (j), the eight metrics associated with the in-flow and out-flow probes were compared by the paired $t$-tests ($N$=24). The results demonstrated that, except for the arriving time, the seven performance metrics associated with the out-flow probes exhibited significant differences in comparison to those associated with the in-flow probes, as reported in Supplementary Table 1. The trend of each metric's difference between in-flow and out-flow probes was the same as their simulation results (Supplementary Fig. 3).

We further computed the correlation between the self-reported flow intensity and the associated performance metrics across all probes and subjects, yielding a total of 288 samples. Prior to analysis, $z$-standardization was applied to normalize the flow intensities and metrics for each subject. The corresponding $t$-tests were also conducted ($N$=288). The results demonstrated that all the eight metrics significantly correlated with the self-reported flow intensity, as reported in Supplementary Table 1. These findings suggest a notable capability of the task performance, as measured by the eight metrics, to differentiate distinct states of flow, thus opening avenues for utilizing these performance metrics to decode variations in flow intensity during experimental sessions.

### 4.3. Flow decoding results

The eight performance metrics were selected to decode the flow variations following the Eq. (4) and (5). For each participant, the flow decoder underwent training, followed by leave-one-out cross-validation, and subsequent statistical testing to assess the efficacy of flow decoding. Cross-validated prediction of the flow intensities of all the probes and all the subjects were depicted against the self-reported flow intensities in the Fig. 6 (a). It indicated a strong and significant correlation between the predicted values and the self-reported flow intensities ($R$=0.81, $P$=1.54×10⁻⁷⁰), suggesting the flow decoder we trained by the performance metrics was highly predictive of the flow probes. At the individual level, Supplementary Fig. 4 exhibited the self-reported and predicted flow intensities of the twelve probes for each participant. The prediction error (measured by $NRMSE$) and the Pearson's $r$ of the correlation between self-reported and predicted flow intensities of twelve probes of each participant were documented in Supplementary Table 2. The results unveiled a significant correlation for 20 participants.

For various subject, different performance metrics were selected to decode flow through the metrics selection process. We calculated the relative decoding contribution of the selected metrics for each participant, as detailed in Supplementary Fig. 5. The summation of the relative decoding contribution attributed to each metric across all the participants was illustrated in Fig. 6 (b). Notably, while the average adjustment rate emerged as the most frequently utilized metric, it was the success rate that yielded the most cumulative contribution to the decoding. Both reaction time and completion time exhibited relatively modest contributions among the eight metrics.

The $P$-value and the probability density distribution of the random-test and the permuted-test for each participant were showcased in Supplementary Table 2, Fig. 6 and 7. The results demonstrated that we were capable of predict the flow intensity by their performance metrics significantly (all the participants passed the random-test and permuted-test, taking $P<0.05$ as the passing criterion). As illustrated in Fig. 6(c), the averaged prediction error (measured by $NRMSE$) of the true flow decoder across participants was both significantly less than that of random-test or permutated-test (true decoder vs. random-test: $t_{24}$=-7.47, $P$=1.37×10⁻⁷, Cohen's $d$=1.52; true decoder vs. permuted-test: $t_{24}$=-7.01, $P$=3.79×10⁻⁷, Cohen's $d$=1.43; compared by paired $t$-tests). There was no significant difference the averaged prediction error of random-test and permuted-test ($t_{24}$=-1.792, $P$=0.086, Cohen's $d$=0.37; compared by paired $t$-test).

The flow intensity was self-reported through the two dimensions of fluency and absorption in the FSS. We repeated the flow decoding process to decode the fluency and absorption during the task using the same performance metrics, taking the subjective responses to the questions belonging to these two dimensions of each flow probe as the ground truth respectively. As listed in Supplementary Table 2 and showed in Fig. 6 (d), the decoding accuracy reduced when the two dimensions were separated (flow intensity vs. fluency: $t_{24}$=-2.88, $P$=0.008, Cohen's $d$=0.59; flow intensity vs. absorption: $t_{24}$=-2.81, $P$=0.01, Cohen's $d$=0.57; fluency vs. absorption: $t_{24}$=-1.57, $P$=0.13, Cohen's $d$=0.32; compared by paired $t$-tests).

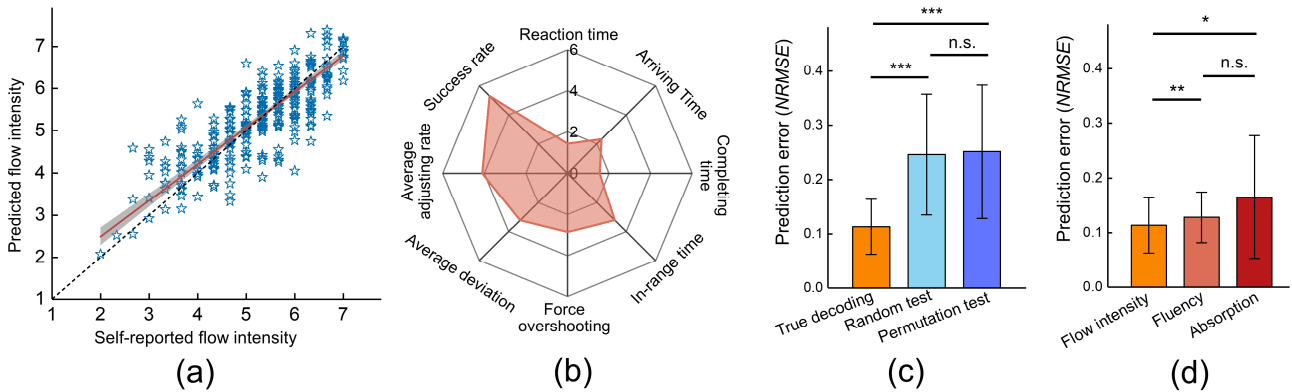

**Fig. 6.** Decoding flow by performance metrics. (a) The correlation between the self-reported and predicted flow intensities of all the probes and all the participants. Each blue star depicts a flow probe and the black dot line denotes the ideal prediction. The red line represents the least squares fit, with grey shading showing the 95% CI. (b) The summation of the relative decoding contribution attributed to each metric across all the participants. (c) The averaged prediction error ($NRMSE$) of true flow decoding, random-test and permutated-test across participants respectively, compared by paired $t$-tests ($N$=24). (d) Prediction errors ($NRMSE$) for decoding flow intensity, fluency, and absorption, compared by paired $t$-tests ($N$=24). *$P<0.05$, **$P<0.01$, ***$P<0.001$, and n.s. stands for no significance. The error bars denote the standard deviation across subjects.

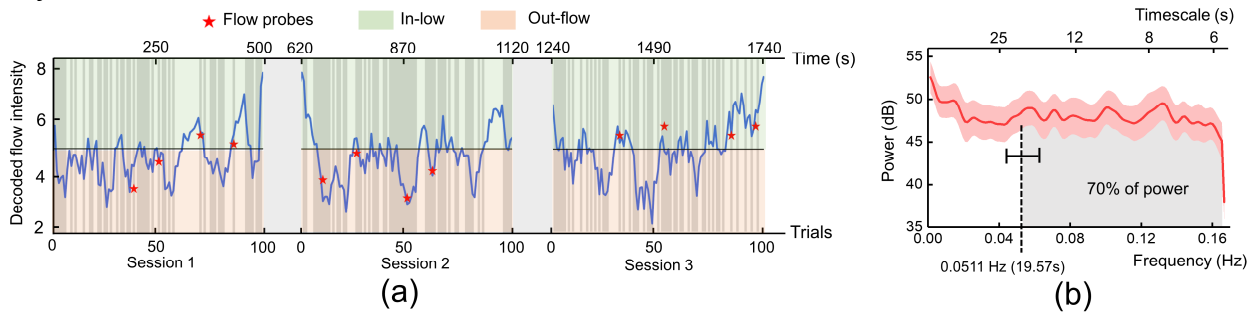

**Fig. 7.** The dynamics of the fluctuation of decoded flow intensity in the time and frequency domain. (a) The time series of the decoded flow intensity (taking example of subject #21). The two grey bands represent the interval between experimental sessions. The median of the self-reported flow intensities of the twelve probes served as the criteria to distinguish in-flow and out-flow states (i.e., the horizonal black line). The vertical semi-transparent grey bands denote the failed trials (average success rate is 47.3%). (b) Estimated PSD of the decoded flow intensity time series. The vertical axis (dB units) is ten times proportional to the log of PSD in original units. The red line depicts the distribution of power on the frequency band from 0 to 0.167Hz (half of the sampling frequency of flow intensity time series 0.333Hz, i.e., inverse of the period of a single trial). The red shadow denotes the standard error of means across participants. Area under the curve over a certain set of frequencies is proportional to how prominent the decoded flow intensity variations are at those frequency band. The horizonal error bar represents the standard deviation of the lower bound of the frequency band in which the 70% of the power occurs across participants.

### 4.4. Dynamics of decoded flow

Based on the validated flow decoder, we were able to obtain the time series of flow intensities during the experiment from the performance metrics of each trial, using the trained flow decoder with the selected decoding-used performance metrics of each subject. An illustrative example was presented in Fig. 7(a), where the decoded flow intensity fluctuated over time but aligned precisely with the self-reporting probes (indicated by the red stars).

Then the PSD analysis was conducted to quantify the timescale of the fluctuation of the decoded flow intensity. As demonstrated in Fig. 7 (b), the results revealed that more than 70% of the power of flow intensity variations occurs at the timescales of 19.57 ± 2.71s (mean ± SD) and faster than that across participants. The results demonstrated that the sub-minute timescale oscillations predominantly contributed to the flow intensity variations.

### V. DISCUSSION

In this work, we validated the feasibility to track the flow variations via task performance on a second timescale, supported by several empirical findings.

First, after extracting eight performance metrics from the fingertip force sequence, we find significant differences in the seven metrics between in-flow and out-flow probes, providing evidence for our hypothesis that different flow states could influence task performance (*H1*). These findings implied that the ambiguity of the relationship between flow and task performance may due to the lacking of explicit behavioral indictors to track the flow.

Second, our findings confirm the hypothesis that flow intensity can be predicted using the task performance metrics (*H2*). The cross-validation and two statistical tests were employed to evaluate the flow decoder and the results demonstrated the high accuracy and robustness of prediction. Compared with the flow decoding studies via physiological or neural data [15][35], we can not only decode flow on a fine timescale, but also consider the flow intensity as a continuous value, not just a state between in-flow and out-flow or among anxiety, flow, and boredom.

Third, we investigated the dynamics of the flow intensity time series. The results highlighted that the flow decoder enabled us to uncover the oscillations of flow in the intervals between self-reporting points. The dynamic nature underlying these internals would have been overlooked without decoding flow using real-time performance metrics. This work demonstrated the ability to overcome the temporal resolution limitations caused by the sparse placement of flow probes, allowing us to investigate the dynamics of flow on a second timescale.

**Inducing flow**. The concept of flow is proposed based on the interviews and observation of the large populations with numerous professions, such as artists and athletes. In the last two decades, more and more experimental studies about flow were conducted in lab settings to investigate the psychological and physiological mechanism underlying flow experience; in many of these studies, computer games were used to induce flow. The difficulty in most computer games, such as falling velocity in Tetris and the keystroke numbers in music games, could be adjusted among several levels, and moderate levels were most commonly chosen as a means of matching average participant skill level [24][25][36]. The insufficient quantification of challenge and skill for these game-based flow induction tasks makes it difficult to achieve customized, precise matching between challenge and skill across different participants.

Flow research began with interviews about daily life experiences before transitioning to self-reporting in ESM methods and eventually to accurate measurements in experimental studies; this demonstrates a trend towards increased explicitness of operational definition and parameterization of challenge and skill. For the $F^3C$ task, the challenge could be explicitly represented by the tolerance range of fingertip force control, which can be continuously altered to match the measured skill. Only under the personally matched challenge, can the human fingertip force control loop be sufficiently activated; this will make the user immersed in the task at hand, merge the action and awareness, and induce the flow experience. Moreover, the temporal resolution and accuracy of fingertip force measurement are essential for this task, supporting the effectiveness of the performance metrics associated with the period of loop and minimum modulation step in the force control model.

**Relationship between flow and motor performance**. Our work involved the construction of a mathematical model aimed at describing the closed-loop force control process. We posited that the influence of flow intensity on performance could be mediated by the updating frequency of the force control loop and the force modulation precision. Although we utilized a simulation to prove that these two key factors could be explicitly represented by the eight performance metrics, but whether and how they play a role within the influence of flow intensity on performance remained unvalidated. In the future we will endeavor to further explore the physiological foundation underlying the $F^3C$ task by neuroscience approaches.

Previous literature has proposed several hypotheses and models discussing the relationship between flow and performance [28]. Since flow is composed of multiple mental components, it is believed that these components mediate the impact of flow on performance, including attention, motivation, motor automaticity, intrinsic rewards, etc. Because the task we used is more concerned with the effect of flow intensity on force control behavior on a sub-minute timescale, this implies that attention plays a more important mediating role [37]. Attention has been verified to have second-level fluctuations, and its impact on task performance variation is explicit [38][39][40]. Behavioral indicators such as reaction time have also been used to characterize attention, inspiring us to decode flow state. However, the mediating role of attention in the F$^3$C task needs to be further verified.

It is worth mentioning that the decoding accuracy reduced when we substitute the ground truth from the flow intensity into its two dimensions in FSS, performance fluency and mental absorption, as demonstrated in Fig. 6(d). These results implied, to an extent, that flow experience cannot be defined simply as attention or motor automaticity. Rather, flow is a complex and nuanced composition of multiple mental components.

**The flow dynamics**. Based on the classical channel model of flow, a series of mathematical models that quantitatively describe the impact of challenge and skill balance on flow were raised [41][42]. These models were initially linear and static. However, the swift transition of flow imply that it is essential to reveal its dynamic characteristics. Researches revealed that the intrinsic nonlinear characteristics such as bifurcation points determine that the flow state will fluctuate indefinitely when challenge and skill are balanced [8][9]. Hence, we proposed that the fluctuation of flow could be decoded at high temporal resolution using task performance.

Previous work related to flow dynamics proposed the Cusp catastrophe model, as well as dynamic properties such as hysteresis and phase transition [5]. The data used for fitting and validating these models usually were derived from the ESM, which limited the development of flow dynamics research. First, the ESM has deficits in measurement accuracy and controlling over irrelevant variables when doing various activities in daily life. Second, the sparsity of ESM sampling limits the timescale of flow dynamics research, and it is hard to reveal the rapid fluctuations of flow through data on fine timescale.

Against this backdrop, the methodology of this work is of great significance for exploring the flow dynamics in the lab. For half a century, the channel model was proposed based on the interviews on diverse professionals, the octuple model was proposed based on ESM data [2], and several cognition models were proposed based on neuroimaging data [37][43][44]. The development of flow theory was accompanied by continuously quantified flow measurement tools and data on a finer timescale. Naturally, the future exploration of flow dynamics also requires measurement methods at even finer timescales, with this work being an important reference.

## VI. Conclusion

Based on the flow-inducing F$^3$C task we proposed, our study showcases the ability to track flow variations through performance metrics on a second timescale. This breakthrough overcomes the temporal limitations of the commonly used self-reporting approaches, like ESM. Our research advances the understanding of the relationship between flow and task performance from a static perspective to a dynamic view. The accuracy of flow decoding is validated by our findings, shedding light on the dynamic nature of flow variations on a second timescale. Our work not only expands the methodological possibilities for investigating flow dynamics with higher temporal resolution but also lays the foundation for future development of closed-loop systems capable of monitoring and regulating flow in the real-world dynamic context, like aerial navigations, benefiting a wider range of individuals.

## Acknowledgement

We thank Yilei Zheng, Zhihao Zhang, Dinghao Xue and Yan Zhang for providing the help on the language, mathematical expressions, and figures. The work is funded by the National Natural Science Foundation of China under Grant No. 61973016.

## Appendix

### A. Adaptive procedure for measuring the force control skill

To achieve a balance between challenge and skill in the F$^3$C task and to induce flow, it was crucial to accurately measure the skill of fingertip force control. Drawing inspiration from the widely-used staircase method in psychophysiology for threshold measurement, we devised an adaptive procedure to measure the skill. The fundamental principle of this procedure was to adjust the task difficulty (represented by the target force range, $\Delta F$) of the next trial based on the performance of the current trial. If the current trial proved too challenging for the subject, the subsequent trial would be made easier; if the current trial was too easy, the next trial would be more difficult. We assumed that as the task difficulty increased, the likelihood of success decreased, and a difficulty level was considered suitable for the subject when the probability of success reached 0.5. If the task difficulty adapted with the performance, it would be converged at the value, under that the possibility of success or failure would be both 0.5. The point of convergence was considered as the measured skill. This converging value was calculated by averaging the task difficulties corresponding to the last ten transition points, where the difficulty sequence reversed, as depicted in Supplementary Fig. 1.

The design of the adaptive procedure considered two key requirements. The first requirement aimed to achieve swift convergence of the task difficulty. This entailed implementing a relatively large adjustment step (i.e., $\Delta F_{i+1}$-$\Delta F_i$, $i$ was the current trial) of difficulty when the task was deemed too easy. The objective was to expedite the convergence process. The second requirement focused on ensuring measurement accuracy. To achieve this, it was essential to minimize the convergence range of the task difficulty and employ a relatively small adjustment step when the difficulty approached the skill. The basic idea was to make the adjustment step of difficulty decrease along with the number of the current trial, as follows:

$$\left| \Delta F_{i+1} - \Delta F_i \right|^{(1)} = \frac{k_1}{i} \qquad (A1)$$

where $k_1$ denoted the coefficient. The sign of $|\Delta F_{i+1} - \Delta F_i|$ would be negative or positive when the current trial was successful or unsuccessful.

Then the performance of the current trial was involved. For simplicity, the completing time of each trial was considered to represent the performance. The adjustment step could be represented as follows:

$$\left| \Delta F_{i+1} - \Delta F_i \right|^{(2)} = \frac{k_2}{t_{Com}} \qquad (A2)$$

where the $t_{Com}$ denoted the completing time of the current trial and $k_2$ denoted the coefficient. The sign of $|\Delta F_{i+1} - \Delta F_i|$ would be negative or positive when the current trial is successful or unsuccessful.

At last, in order to avoid the adjustment step being too large relative to the current difficulty, we set a limit of the adjustment step by the current difficulty, which could be represented as follows:

$$\left| \Delta F_{i+1} - \Delta F_i \right|^{(3)} = 0.5 \Delta F_i \qquad (A3)$$

Combine these factors, the adjustment step of difficulty was determined by the number, performance, and the difficulty of the current trial, which was derived as:

$$\left| \Delta F_{i+1} - \Delta F_i \right| = \min \left\{ \frac{k_1}{i}, \frac{k_2}{t_{Com}}, 0.5 \Delta F_i \right\} \qquad (A4)$$

The adjustment step was the minimum among Eq. (A1), (A2), and (A3). The sign of $|\Delta F_{i+1} - \Delta F_i|$ would be negative or positive when the current trial was successful or unsuccessful.

*B. Mathematical model on fingertip force control process*

We established the mathematical model to describe the closed-loop fingertip force control process, as depicted in Fig. 2(b), following the physiological fundamentals. During the task, the fingertip force modulation could be represented as follows:

$$F^{(t+\Delta T)} = F^{(t)} + \Delta F_M^{(t)} + w_F \qquad (A5)$$

where $F^{(t+\Delta T)}$ and $F^{(t)}$ denoted the pressing force at the timepoint $t+\Delta T$ and $t$ respectively. $\Delta T$ denoted the period of the force control loop and $w_F$ denoted the force output noise. $\Delta F_M^{(t)}$ represented the modulated force, which was determined by the motor command $\Delta F_C^{(t)}$ and limited by the minimum modulation step of fingertip force $\delta F$, as follows:

$$\Delta F_M^{(t)} = \begin{cases} \left[ \dfrac{\Delta F_C^{(t)}}{\delta F} \right] \delta F, & \left| \Delta F_C^{(t)} \right| > \delta F \\ \delta F, & \left| \Delta F_C^{(t)} \right| \le \delta F \end{cases} \qquad (A6)$$

The motor command was generated after motor decision making, in which the participants make decisions about the fingertip force modulation based on the visual information. The motor command could be derived as follows:

$$\Delta F_C^{(t)} = k_F \left( H_0 - H^{(t)} \right) + w_C \qquad (A7)$$

where $H^{(t)}$ and $H_0$ represented the current height and the target height observed by the subject. $w_C$ denoted the error of the decision making and $k_F$ is the coefficient. Eq. (A7) showed that the human update the motor command based on the bias between the current and the target height.

The current fingertip force was measured by the force transducer and transformed into the height of the virtual disk, which could be represented as follows:

$$H^{(t)} = k_H \left( F^{(t)} + w_M \right) + w_V \qquad (A8)$$

where $w_M$ and $w_V$ respectively denoted the error of the fingertip force measurement and subject's visual observation. Coefficient $k_H$ is the inverse of $k_F$. The force control model for the F³C task encompassed the Eq. (A5) to (A8), as depicted in Supplementary Fig. 2, where the $\Delta T$ and $\delta F$ would be influenced by the flow intensity.

Based on this model, we simulated the closed-loop force control process and investigate how the performance metrics are influenced by $\Delta T$ and $\delta F$. The results showed that all the eight performance metrics were capable of indicating the influence of $\Delta T$ and $\delta F$ (Supplementary Fig. 3).

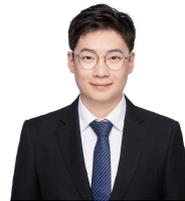
**Bohao Tian** received the B.E. degree in mechanical engineering from Beihang University, Beijing, China, in 2019. Currently, he is working toward a Ph.D. in School of Mechanical Engineering and Automation at Beihang University, Beijing, China. His research interests include cognitive neuroscience, haptic interaction, and brain-computer interactions.

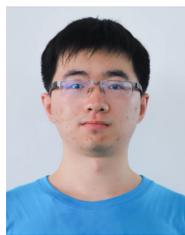
**Shijun Zhang** received the B.E. degree in mechanical engineering from Beihang University, Beijing, China, in 2022. Currently, he is working toward a M.E. in School of Mechanical Engineering and Automation at Beihang University, Beijing, China. His research interests include cognitive neuroscience, haptic interaction, and brain-computer interface.

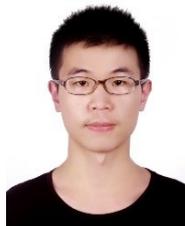
**Sirui Chen** is an undergraduate student in School of Mechanical Engineering and Automation at Beihang University, Beijing, China. His research interests include haptic interaction and robotics.

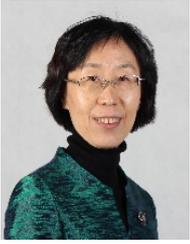

**Yuru Zhang** (M'95-SM'08) received the Ph.D. degrees in mechanical engineering from Beihang University, Beijing, China in 1987. Currently she is a professor at the Division of Human-Machine Interaction at the State Key Laboratory of Virtual Reality Technology and System, Beihang University. Her research interests include haptic human-machine interface, medical robotic system, robotic dexterous manipulation, and virtual prototyping. She is a senior member of IEEE, and a member of ASME.

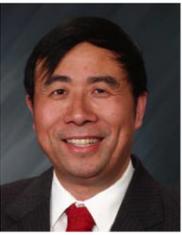

Kaiping Peng received a Ph.D. degree in Social Psychology from the University of Michigan, Ann Arbor, USA in 1997. Currently he is a Professor at the Department of Psychology, School of Social Sciences, Tsinghua University, Beijing, China. His research interests include positive psychology, cross-cultural psychology and social psychology.

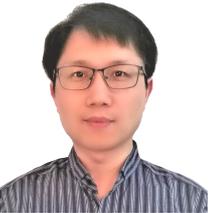

Hongxing Zhang received a Ph.D. degree in biochemistry and molecular biology from Institute of Biophysics of Chinese Academy of Sciences, Beijing, China in 2008. Currently he is a Professor at the State Key Laboratory of Proteomics, Beijing Proteome Research Center, National Center for Protein Sciences (Beijing), and Beijing Institute of Life Omics. His research interests include Genetics and genomics, gene function and molecular mechanism, bioinformatics, and AI analysis tools.

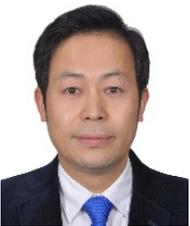

**Dangxiao Wang** (M'05-SM'13) received a Ph.D. degree in mechanical engineering from Beihang University, Beijing, China in 2004. Currently he is a Professor at the State Key Laboratory of Virtual Reality Technology and Systems in Beihang University. His research interests include haptic device, haptic rendering, cognitive-haptics and medical robotic systems. He is a senior member of IEEE. He had been the chair of Executive Committee of the IEEE Technical Committee on Haptics (IEEE TCH) from 2014 to 2017.

# Supplementary Materials

**Supplementary Table 1.** The statistical results of the eight performance metrics.

| Performance metric | Paired $t$-test ($N$=24) | | | Correlation ($N$=288) | |
|---|---|---|---|---|---|
| | $t$-value | $P$-value | Cohen's $d$ | Pearson's $r$ | $P$-value |
| Reaction time | -2.74 | 0.012* | 0.56 | -0.25 | $1.23 \times 10^{-5}$*** |
| Arriving time | -1.85 | 0.078 | 0.38 | -0.12 | 0.044* |
| Completing time | -5.40 | $1.76 \times 10^{-5}$*** | 1.10 | -0.28 | $2.04 \times 10^{-6}$*** |
| In-range time | 4.94 | $5.39 \times 10^{-5}$*** | 1.01 | 0.30 | $2.53 \times 10^{-7}$*** |
| Force overshooting | -3.06 | 0.006** | 0.62 | -0.27 | $3.80 \times 10^{-6}$*** |
| Average deviation | -2.67 | 0.014* | 0.54 | -0.30 | $2.18 \times 10^{-7}$*** |
| Average adjusting rate | -3.12 | 0.005** | 0.64 | -0.24 | $4.16 \times 10^{-5}$*** |
| Success rate | 6.24 | $2.3 \times 10^{-6}$*** | 1.27 | 0.39 | $9.44 \times 10^{-12}$*** |

Note. *$P$<0.05, **$P$<0.01, and ***$P$<0.001. Each performance metric of in-flow probes and out-flow probes was compared by the paired $t$-tests ($N$=24). The Pearson correlations between the self-reported flow intensity and each associated performance metrics across all probes and subjects were conducted ($N$=288).

**Supplementary Table 2.** Flow decoding results for each subject.

| Subject | Flow intensity prediction error | Pearson's $r$ (self-reported vs. decoded flow) | $P$-value of random-test | $P$-value of permuted-test | Fluency prediction error | Absorption prediction error |
|---|---|---|---|---|---|---|
| 1 | 13.38% | 0.57 | 0.007** | 0.018* | 19.05% | 10.28% |
| 2 | 7.30% | 0.76** | 0.003** | 0.003** | 8.41% | 5.61% |
| 3 | 13.05% | 0.41 | 0.038* | 0.033* | 15.39% | 19.46% |
| 4 | 6.85% | 0.76** | 0.001** | 0.001** | 11.04% | 14.62% |
| 5 | 20.12% | 0.74** | <0.001*** | 0.004** | 22.89% | 21.78% |
| 6 | 7.38% | 0.66* | 0.001** | 0.017* | 7.20% | 11.10% |
| 7 | 4.25% | 0.94*** | <0.001*** | <0.001*** | 12.56% | 33.97% |
| 8 | 21.71% | 0.64* | 0.007** | 0.012* | 12.90% | 56.79% |
| 9 | 15.43% | 0.77** | 0.002** | 0.007** | 16.93% | 16.44% |
| 10 | 20.34% | 0.81** | <0.001*** | 0.001** | 12.97% | 18.96% |
| 11 | 9.72% | 0.59* | 0.002** | 0.002** | 13.87% | 4.41% |
| 12 | 7.95% | 0.81** | <0.001*** | <0.001*** | 9.28% | 8.68% |
| 13 | 9.95% | 0.67* | 0.005** | 0.003** | 10.58% | 11.65% |
| 14 | 5.21% | 0.73** | 0.002** | 0.001** | 3.36% | 16.19% |
| 15 | 16.71% | 0.52 | 0.016* | 0.015* | 17.30% | 21.62% |
| 16 | 6.42% | 0.82** | <0.001*** | <0.001*** | 7.64% | 10.02% |
| 17 | 13.59% | 0.67* | <0.001*** | <0.001*** | 15.86% | 18.49% |
| 18 | 10.70% | 0.7* | <0.001*** | 0.001** | 13.22% | 9.23% |
| 19 | 6.31% | 0.93*** | <0.001*** | <0.001*** | 13.52% | 16.73% |
| 20 | 9.07% | 0.87*** | <0.001*** | <0.001*** | 7.55% | 12.51% |
| 21 | 12.13% | 0.67* | <0.001*** | <0.001*** | 16.68% | 24.13% |
| 22 | 17.15% | 0.43 | 0.007** | 0.007** | 16.83% | 20.10% |
| 23 | 13.34% | 0.58* | 0.004** | 0.003** | 16.27% | 12.88% |
| 24 | 4.23% | 0.62* | 0.001** | 0.001** | 6.48% | 0.00% |
| Mean | 11.34% | ----- | ----- | ----- | 12.82% | 16.49% |

Note. *$P$<0.05, **$P$<0.01, and ***$P$<0.001. The prediction error for each participant is measured by the normalized root mean square error (*NRMSE*). The third column shows the Pearson' $r$ of the correlation between the self-reported and the decoded flow intensities of twelve flow probes for each participant. The asterisks in the third column indicate the correspding $t$-tests results.

**Supplementary Fig. 1.** Schematic of force control skill measurement approach, taking subject#24 as an instance. Each blue dot denotes the task difficulty (i.e., target force range, $\Delta F$) in each trial during the measurement session (consisting of 50 trials). The red dots in subfigure represent the last ten transition points, where the task difficulty sequence reverse, and the red line indicates the measured skill by averaging the task difficulties corresponding to the transition points.

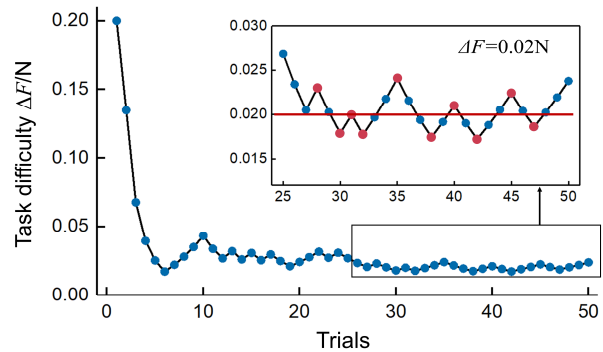

**Supplementary Fig. 2.** The mathematical model describing the fingertip force control process during the F³C task. The increasing of flow intensity would expedite the force control loop and lead to the decreasing of time delay $\Delta T$ and mininum modulation setp $\delta F$, resulting in the change of force control performance.

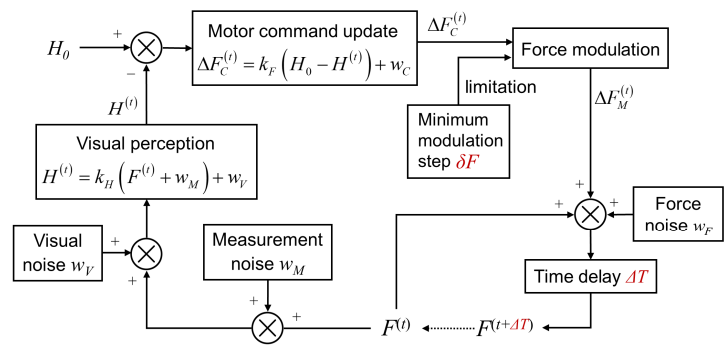

**Supplementary Fig. 3.** The simulation results of the force control model. (a) The example of simulated fingertip force sequence under in-flow and out-flow states. The dot lines denote the current target range of force control, $\Delta F$=0.055N. The simulation was implemented following the mathmatical model of force control process (see Appendix B and Supplementary Fig. 2), i.e., the Eq. (A5) to (A8). According to our hypothesis, the force control loop would be facilitated to work along with the increase of flow intensity, and $\Delta T$ and $\delta F$ would decrease in the meantime. The manipulation of the two different flow state is conducted by altering the $\Delta T$ (in-flow: 0.15s vs. out-flow: 0.3s) and $\delta F$ (in-flow: 0.015N vs. out-flow: 0.03N) in the mathematical model of force control process (see Appendix B), without changing any other parameters in the model. (b) to (g) show the six performance metrics extracted from the simulated force output sequence under out-flow and in-flow states. The simulation and metrics calculation were conducted for 100 times, the same as the number of trials in a single session. The metric reaction time can be derived directly from the force control model (i.e., reaction time=$\Delta T$). The metric success rate cannot be calculated for a single trial (the average success rate: in-flow is 0.61 and out-flow is 0.47), so the results of the remaining six metrics are demonstrated. The metrics under out-flow and in-flow states are compared by the two-tailed paired-samples $t$-tests ($N$=100). Hinges of boxplots represent the first and third quartile and whiskers span from the smallest to the largest value of the data. *$P$<0.05, **$P$<0.01, and ***$P$<0.001.

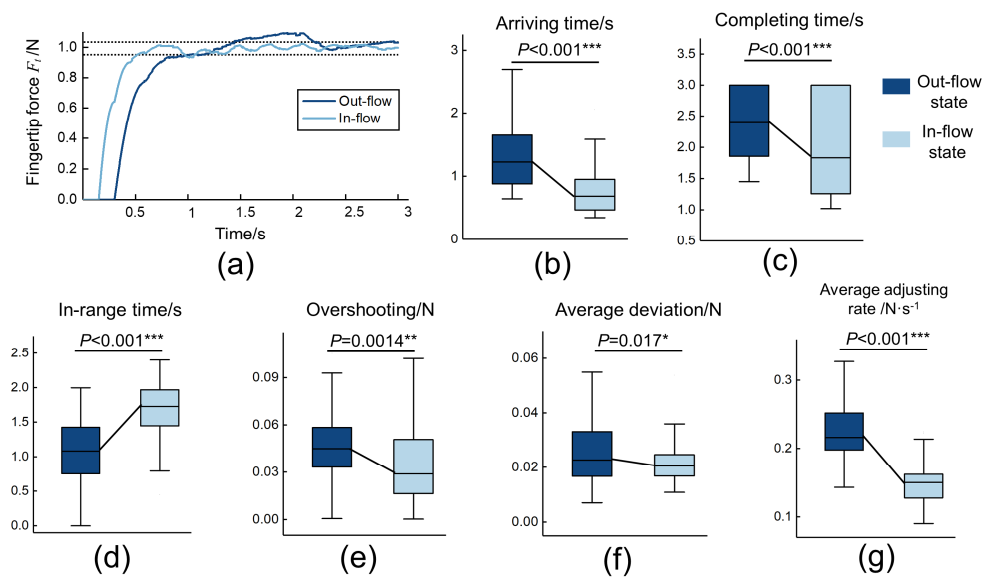

**Supplementary Fig. 4**. Cross-validated decoding performance of each participant. The prediction of flow intensity of each participant was obtained following the decoding framework, compared with the self-reporting flow intensity in each probe. Each blue dot represents a flow probe and the grey lines indicate the ideal prediction.

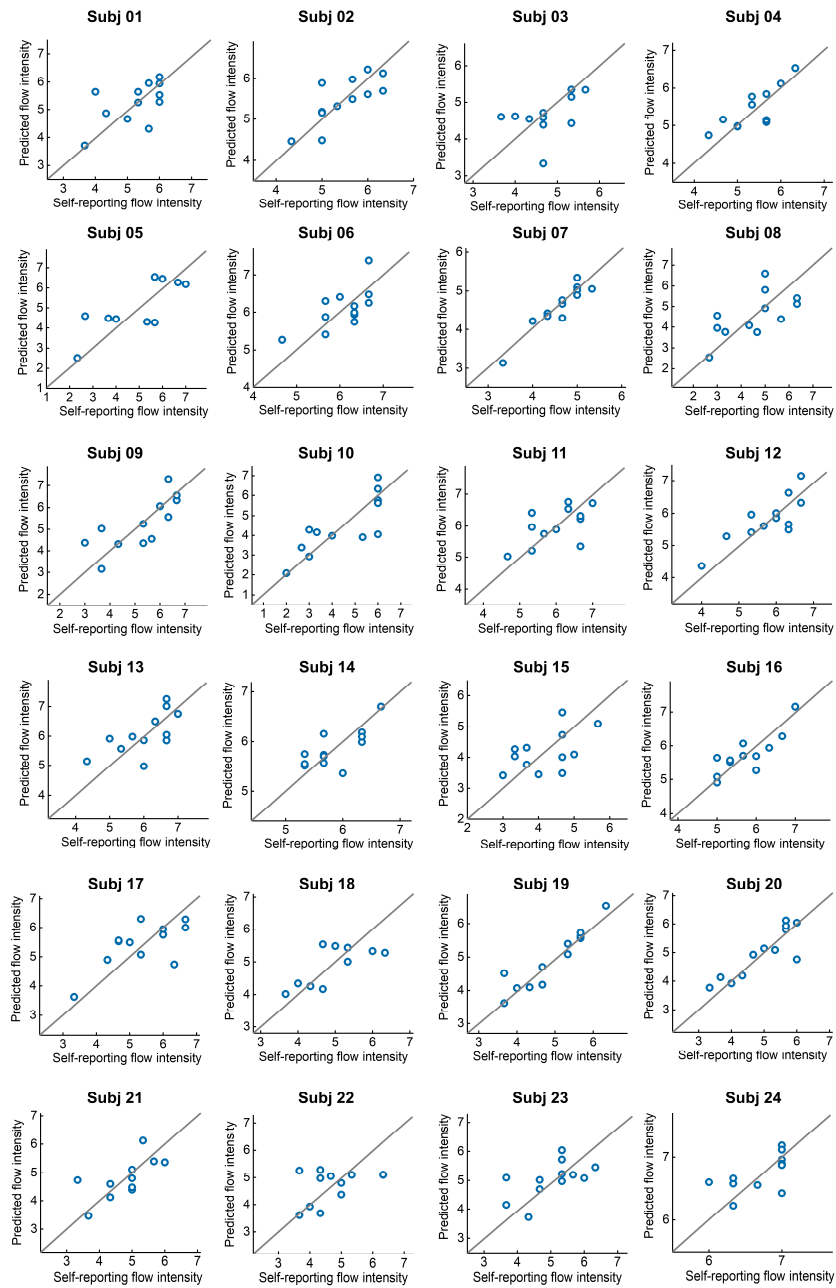

**Supplementary Fig. 5.** The selection and relative decoding contribution of decoding-used performance metrics for each subject. Note. In the Eq. (4), the coefficients $t_1$, $t_2$..., $t_n$ in vetor $\boldsymbol{T}^*$ are capable of measuring the decdoding contributions of their corresponded metrcis after the decoding-used metrics are $z$-standardized for each participant. The relative decoding contribution of a metric refers to the proportion of the absolute value of its corrsponding coefficient $t$ to the sum of the absolute values of all the coefficients within $\boldsymbol{T}$ for a specific subject. Consequently, for each subject, the summation of the relative decoding contributions of all the decoding-used metrics amounts to 1. The null positions within the table indicate that the corresponding metrics are not selected for decoding.

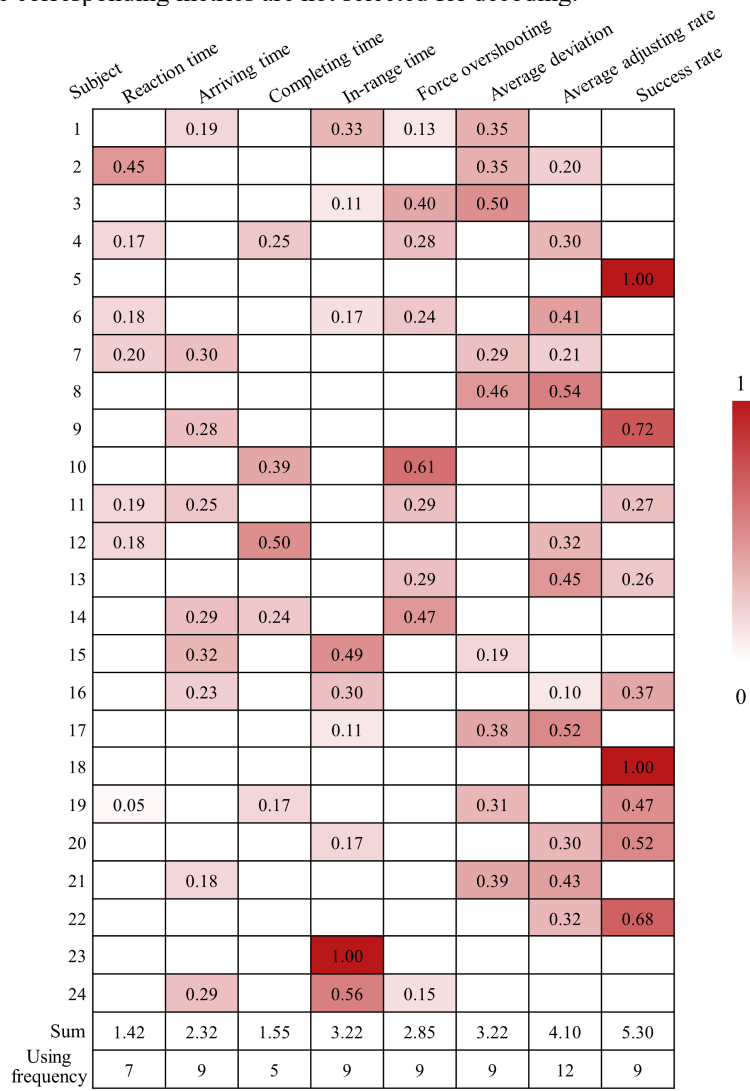

| Subject | Reaction time | Arriving time | Completing time | In-range time | Force overshooting | Average deviation | Average adjusting rate | Success rate |
|---|---|---|---|---|---|---|---|---|
| 1 | | 0.19 | | 0.33 | 0.13 | 0.35 | | |
| 2 | 0.45 | | | | | 0.35 | 0.20 | |
| 3 | | | | 0.11 | 0.40 | 0.50 | | |
| 4 | 0.17 | | 0.25 | | 0.28 | | 0.30 | |
| 5 | | | | | | | | 1.00 |
| 6 | 0.18 | | | 0.17 | 0.24 | | 0.41 | |
| 7 | 0.20 | 0.30 | | | | 0.29 | 0.21 | |
| 8 | | | | | | 0.46 | 0.54 | |
| 9 | | 0.28 | | | | | | 0.72 |
| 10 | | | 0.39 | | 0.61 | | | |
| 11 | 0.19 | 0.25 | | | 0.29 | | | 0.27 |
| 12 | 0.18 | | 0.50 | | | | 0.32 | |
| 13 | | | | | 0.29 | | 0.45 | 0.26 |
| 14 | | 0.29 | 0.24 | | 0.47 | | | |
| 15 | | 0.32 | | 0.49 | | 0.19 | | |
| 16 | | 0.23 | | 0.30 | | | 0.10 | 0.37 |
| 17 | | | | 0.11 | | 0.38 | 0.52 | |
| 18 | | | | | | | | 1.00 |
| 19 | 0.05 | | 0.17 | | | 0.31 | | 0.47 |
| 20 | | | | 0.17 | | | 0.30 | 0.52 |
| 21 | | 0.18 | | | | 0.39 | 0.43 | |
| 22 | | | | | | | 0.32 | 0.68 |
| 23 | | | | 1.00 | | | | |
| 24 | | 0.29 | | 0.56 | 0.15 | | | |
| Sum | 1.42 | 2.32 | 1.55 | 3.22 | 2.85 | 3.22 | 4.10 | 5.30 |
| Using frequency | 7 | 9 | 5 | 9 | 9 | 9 | 12 | 9 |

**Supplementary Fig. 6.** Probability density distribution of the prediction errors (measured by the normalized root mean square error, *NRMSE*) for 1,000 random sets of the flow intensity of each participant, compared with the true set of the flow intensity in each probe. The orange line denotes the prediction error in the decoding taking true set of the flow intensity in each probe.

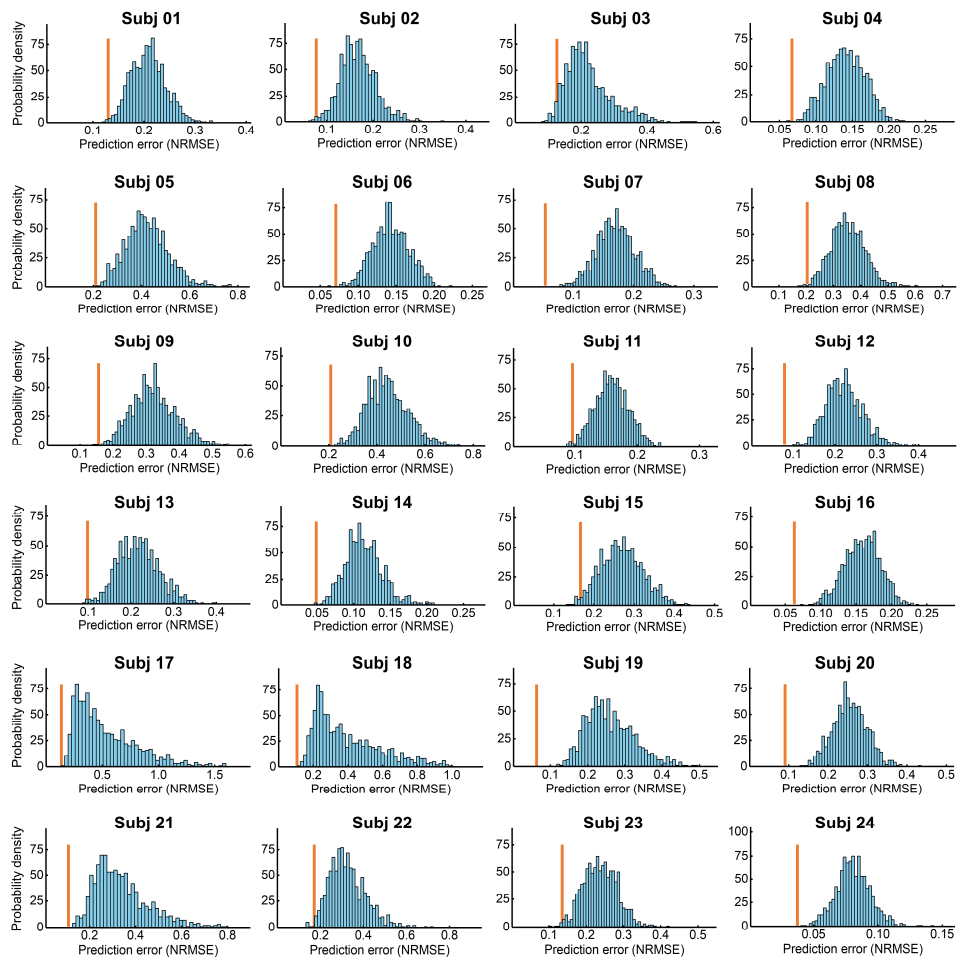

**Supplementary Fig. 7.** Probability density distribution of the prediction errors (measured by the normalized root mean square error, *NRMSE*) for 1,000 permutated sets of the flow intensity of each participant, compared with the true set of the flow intensity in each probe. The orange line denotes the prediction error in the decoding taking true set of the flow intensity in each probe.

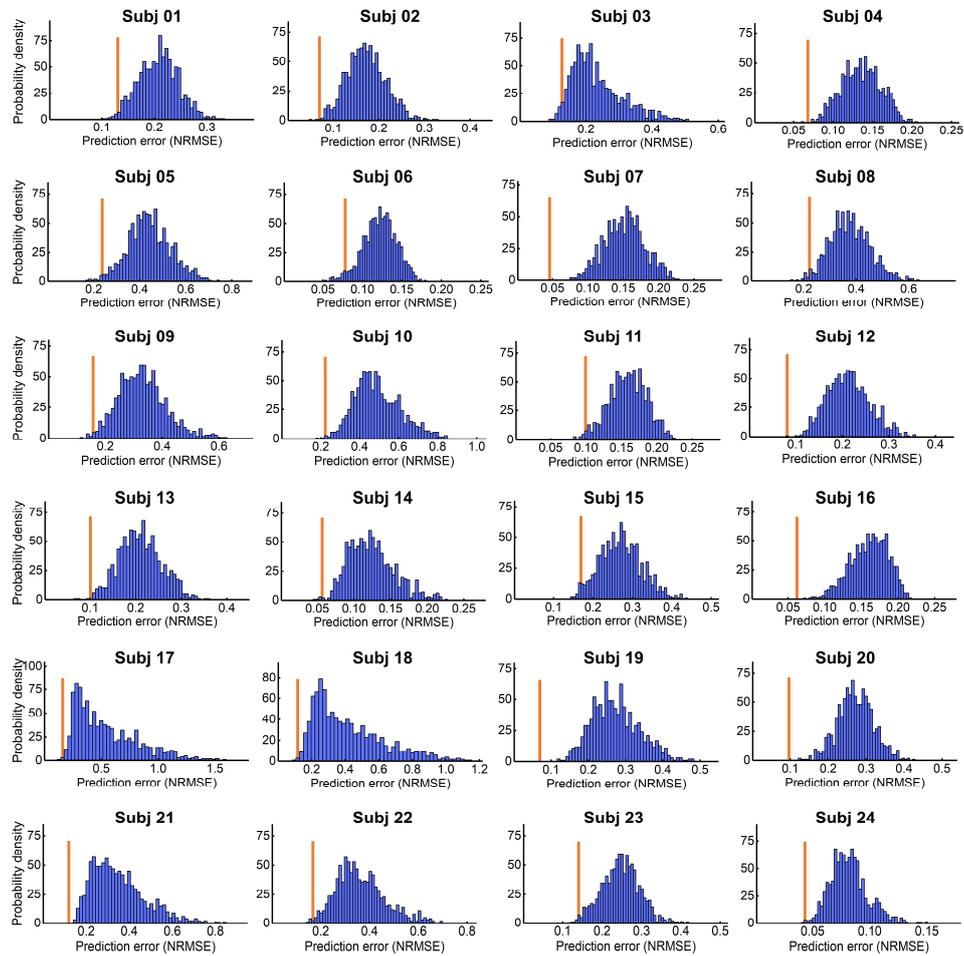